\newif\ifsubmission
\submissionfalse

\ifsubmission
  \documentclass[aps,prl,preprint,superscriptaddress,nofootinbib]{revtex4-2}
\else
  \documentclass[aps,prl,twocolumn,superscriptaddress,reprint,nofootinbib]{revtex4-2}
\fi

\usepackage{graphicx}
\ifsubmission\usepackage{lineno}\linenumbers\fi
\usepackage{microtype}%
\usepackage{dcolumn}%
\usepackage{booktabs}%
\usepackage[utf8]{inputenc}
\usepackage[hidelinks]{hyperref}
\usepackage{amsmath}
\usepackage{siunitx}
\DeclareSIUnit{\atom}{atom}
\DeclareSIUnit{\angstrom}{\text{\AA}}%
\usepackage{cleveref}
\usepackage[version=3]{mhchem}
\usepackage{xcolor}
\graphicspath{{figures/}}
\begin{document}

\title{Structural Chirality from Short-Range Order
in Heteroanionic Materials}
\author{Benjamin J. Morgan}
\affiliation{Department of Chemistry, University of Bath, Claverton Down, Bath, BA2 7AY, United Kingdom}

\date{\today}

\begin{abstract}
Established routes to structural chirality in inorganic crystals predominantly depend on symmetry-lowering atomic displacements of an achiral parent structure.
We show that chirality can instead be forced by the short-range order of two anion species over the sites of an achiral parent, independent of atomic displacements.
\ce{ReO3}-type oxyfluorides with a 2:1 anion stoichiometry, such as \ce{NbO2F}, exhibit two ordering patterns, cis octahedral coordination and period-three anion-chain ordering, both of which are rooted in the off-centring of $d^0$ cations.
By direct enumeration, we prove that every arrangement combining both orderings on the smallest commensurate cell is chiral, belonging to a Sohncke space group.
For \ce{NbO2F}, density-functional theory (DFT) calculations predict a homochiral $P3_121$ ground state.
Wang--Landau Monte Carlo simulations of a DFT-trained cluster-expansion model predict that the chiral phase remains the equilibrium state up to a first-order transition at \qty{494}{\kelvin}.
We also identify three known vacancy-ordered compounds, \ce{BaZnO2}, \ce{BaCoO2} and $\beta$-\ce{Fe2SeO}, whose structures each map onto one of the enumerated chiral arrangements.
Together, these results establish short-range order as a route to structural chirality, and mark heteroanionic and vacancy-ordered materials in which both orderings are active as candidates for new chiral phases.
\end{abstract}

\maketitle

\section{Introduction}

Structural chirality---the absence of improper symmetry operations in a crystal's space group, rendering the structure non-superimposable on its mirror image---gives rise to a range of functional phenomena in inorganic solids, including optical activity~\cite{GlazerAndStadnicka_JApplCrystallogr1986}, magnetochiral anisotropy, chirality-induced spin selectivity, and chiral phonons~\cite{BousquetEtAl_JPhysCondensMatter2025}.

Established routes to structural chirality in inorganic crystals predominantly depend on displacive symmetry breaking~\cite{BousquetEtAl_JPhysCondensMatter2025,GomezOrtizEtAl_JApplCrystallogr2026,GomezOrtizEtAl_PhysRevB2025}.
In the simplest case, an unstable phonon mode of a high-symmetry achiral parent condenses and displaces the atoms off their high-symmetry positions into a chiral structure.
Chirality can also arise from configurational ordering, in which distinct chemical species (or vacancies) adopt an ordered arrangement over the sites of a structure.
In some materials, the ordering is coupled to displacements: in oxyfluoride double perovskites, the combination of octahedral tilting and anion ordering has been analysed as producing chiral and chiral-polar structures~\cite{CharlesEtAl_ChemMater2018}.
In other cases, configurational ordering alone gives chiral structures: for example, cation ordering in inverse spinels such as \ce{Zn2TiO4}~\cite{StevanovicEtAl_JAmChemSoc2011,LiuEtAl_InorgChem2019} and vacancy ordering in 3C-\ce{Fe7Se8}~\cite{PariseEtAl_ActaCrystallogrB1979}.

Here we show that structural chirality can be forced by the combination of two chemically driven short-range orderings.
For a \ce{ReO3}-type parent lattice with two anion species in a 2:1 ratio, cis coordination at each cation and period-three order along every $\langle001\rangle$ anion chain allow only twelve symmetry-inequivalent arrangements on the smallest commensurate cell ($3\times3\times3$), and all twelve are chiral.
Both orderings are observed in \ce{NbO2F} and related $d^0$ oxyfluorides with the \ce{ReO3}-type structure, where they are rooted in the off-centring of the cations~\cite{BrinkEtAl_JSolidStateChem2002,WithersEtAl_Polyhedron2007,DabachiEtAl_InorgChem2017,LegeinEtAl_JAmChemSoc2024}.
Each ordering alone allows achiral arrangements, so chirality is forced only when both are present.
In any larger structure that combines both orderings, every $3\times3\times3$ region matches one of these twelve arrangements, so the orderings force chirality locally at any size.

For \ce{NbO2F}, density-functional theory (DFT) calculations show that both orderings are energetically favoured and that twin boundaries between regions of opposite hand cost energy, so the predicted ground state is homochiral.
This ground state is the $P3_121$ arrangement, in which Nb--F--Nb bonds wind in helices along $\langle111\rangle$.
The preference for $P3_121$ can be understood as electrostatic in origin, as $P3_121$ maximises the separation between fluorides on adjacent parallel chains.
Replica-exchange Wang--Landau Monte Carlo simulations of a DFT-trained cluster-expansion model predict that the chiral phase remains the equilibrium state up to a first-order transition at $T_\mathrm{c} = \qty{494}{\kelvin}$.
Both orderings persist above $T_\mathrm{c}$, although at finite temperature neither holds exactly.
The transition instead orders the chains relative to one another: above $T_\mathrm{c}$, the period-three order on different chains has no fixed relative phase, and below it the chains lock into the helical arrangement of the ground state.

The result that cis coordination and period-three order together force chirality holds for any system with a \ce{ReO3}-type network in which both short-range orderings are active.
Further DFT calculations for \ce{TaO2F}, which shares the structure and $d^0$ chemistry of \ce{NbO2F}, predict the same homochiral $P3_121$ ground state.
We also identify three known vacancy-ordered compounds, \ce{BaZnO2}~\cite{vonSchneringEtAl_ZAnorgAllgChem1960}, \ce{BaCoO2}~\cite{Spitsbergen_ActaCrystallogr1960,DiattaEtAl_JAmChemSoc2024} and $\beta$-\ce{Fe2SeO}~\cite{ValldorEtAl_AngewChemIntEd2016,ShamovaEtAl_ChemMater2026}, whose structures each map onto one of the twelve chiral arrangements, and whose chirality can therefore be understood in terms of the two short-range orderings.
Together, these results establish short-range order as a route to structural chirality, and mark heteroanionic and vacancy-ordered materials in which both orderings are active as candidates for new chiral phases.

\section{Short-range order in \ce{ReO3}-type \ce{MX2Y} materials}
\label{sec:rules}

\begin{figure}[tb]
  \centering
  \resizebox{6.5cm}{!}{\includegraphics*{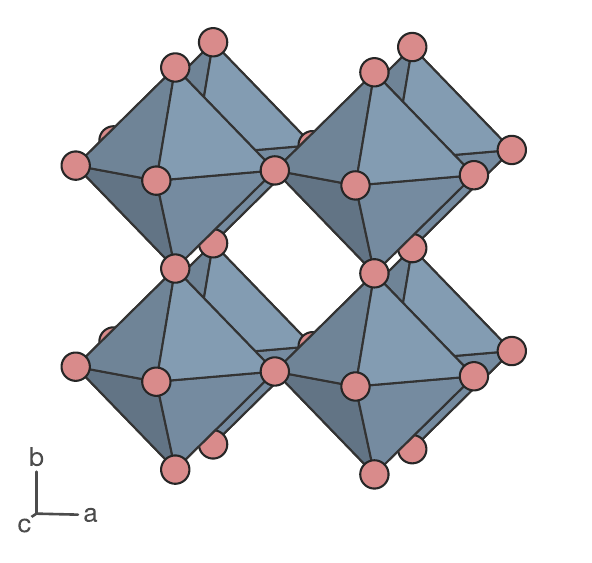}}%
    \caption{\label{fig:ReO3}
    The achiral \ce{ReO3}-type structure, comprising a cubic $Pm\bar{3}m$ framework of corner-sharing \ce{MX6} octahedra.
    }
\end{figure}

\ce{ReO3}-type mixed-anion materials adopt a simple cubic ($Pm\bar{3}m$) framework of corner-sharing \ce{MX6} octahedra (Fig.~\ref{fig:ReO3}), with the anion species jointly occupying a single crystallographic site with no long-range order~\cite{FrevelAndRinn_ActaCrystallogr1956}. In materials with $d^0$ cations, such as \ce{NbO2F}, \ce{TaO2F} and \ce{TiOF2}, cations with polar coordination can move off-centre, giving, for example, shorter \ce{M-O} and longer \ce{M-F} bonds. This off-centring increases the net local bonding~\cite{KunzAndBrown1995,HalasyamaniAndPoeppelmeier1998,WithersEtAl_Polyhedron2007}, making polar coordination energetically preferred over non-polar coordination.

In \ce{NbO2F}, this preference for polar coordination drives two short-range orderings. Under local charge neutrality~\cite{Pauling1929}, every niobium is coordinated by four oxides and two fluorides. The two fluorides preferentially sit cis rather than trans, on adjacent rather than opposite octahedral vertices, since only a cis arrangement leaves the niobium free to move off-centre.

A second ordering extends along the $\langle001\rangle$ anion chains running along each cube axis. Fluoride ions preferentially occupy every third site, giving a period-three oxide-oxide-fluoride (OOF) pattern, observed as $\{hk1/3\}^{*}$ diffuse scattering in electron diffraction~\cite{BrinkEtAl_JSolidStateChem2002} and consistent with solid-state NMR~\cite{DabachiEtAl_InorgChem2017}. This ordering avoids collinear \ce{F-Nb-F} pairs, which are incompatible with cation off-centring along the chain direction~\cite{WithersEtAl_Polyhedron2007}.
\section{The combinatorial theorem}
\label{sec:theorem}

\begin{figure*}[tb]
  \centering
  \resizebox{\textwidth}{!}{\includegraphics*{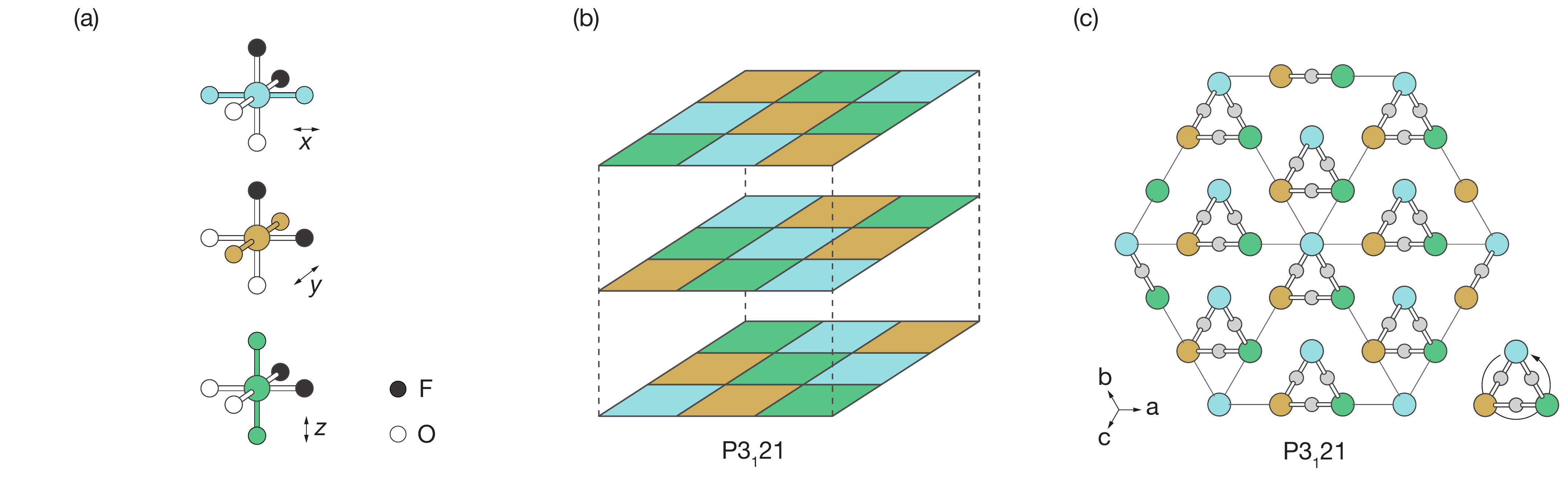}}%
    \caption{\label{fig:p3121}
    (a)~The labelling rule used for structure enumeration, shown for the three axes: at each cation the two fluorides (black) sit cis, leaving one octahedral axis where both are oxides (coloured). That fluoride-free axis ($x$, $y$ or $z$) sets the cation label. (b)~Cation labelling for the $P3_121$ ground state: the cations of the $3\times3\times3$ cell are coloured by their fluoride-free axis. (c)~The same $P3_121$ configuration, viewed down $\langle111\rangle$ with only the \ce{Nb-F} bonds drawn. Each small triangle is one Nb--F--Nb helix seen end-on. The fluoride-free axis cycles $x\to y\to z$ out of the page (inset arrow), giving the $3_1$ screw about this $\langle111\rangle$ axis.
    }
\end{figure*}

We now ask which anion arrangements on a \ce{ReO3}-type anion sublattice satisfy both cis-\ce{NbO4F2} coordination at every cation and period-three fluoride ordering along every $\langle001\rangle$ anion chain. The smallest supercell that accommodates the period-three ordering is the $3\times3\times3$ supercell, with 27 cation sites and 81 anion sites. We enumerate every anion configuration on this supercell that satisfies both rules, and classify the results by symmetry.

Direct enumeration over all anion configurations is intractable: placing the 27 fluorides over the 81 anion sites gives $\sim\!\!10^{21}$ total configurations. Instead, we use the constraint that, for allowed configurations, every cation is cis-coordinated, which defines a smaller combinatorial space that can be efficiently searched. A cation with cis-\ce{O4F2} coordination has anion pairs (O,F), (O,F) and (O,O) along its three $\langle001\rangle$ axes, so exactly one octahedral axis is fluoride-free. Each cis cation can therefore be represented by a single label describing its fluoride-free axis (Fig.~\ref{fig:p3121}a). OOF ordering then requires each chain to contain exactly one cation whose fluoride-free axis lies along the chain direction (Supplementary Information). To enumerate all cis + OOF configurations, we perform a depth-first search over the cis labelling space. As each $\langle001\rangle$ chain is completed, we count its along-chain labels. If the count is not one, the labelling does not describe a cis + OOF configuration: we prune that branch and backtrack. The valid labellings correspond one-to-one with the cis + OOF configurations. They number \num{10752}, and under the full parent symmetry (the 48 point-group operations of $Pm\bar{3}m$ and the 27 lattice translations of the supercell) they reduce to twelve symmetry-inequivalent orbits.

Every one of these twelve orbits is chiral. For each orbit we compute the stabiliser, the set of operations that map a representative configuration onto itself. Every stabiliser consists solely of proper rotations, with no mirror, inversion, or rotoinversion. The orbits therefore belong to Sohncke space groups: one $P3_121$ (stabiliser of order 54), one $P3_1$ (order 9), three $C2$ (orders 2, 2 and 6) and seven $P1$.

Neither of the two ordering rules produces chirality on its own. OOF ordering alone leaves mirror-symmetric arrangements available: a mirror perpendicular to a chain through its fluoride maps such an arrangement onto itself. Cis coordination alone likewise leaves arrangements with mirrors that exchange the two fluorides at a cation, for example when fluorides and oxides alternate along the $x$ and $y$ chains and the $z$ chains are all oxide. Imposing both rules together, however, leaves only the twelve chiral orbits on the $3\times3\times3$ cell.

The chirality forced by the two rules also survives any lowering of the parent symmetry to a subgroup of $Pm\bar{3}m$. Intersecting a proper-rotation stabiliser with a subgroup leaves only proper rotations. A displacive distortion of the \ce{ReO3}-type framework lowers its symmetry to such a subgroup, and so cannot make a chiral configuration achiral. The chirality is therefore a property of the anion ordering alone, independent of the displacive state of the framework.

On cells larger than $3\times3\times3$, arrangements that obey both rules can be achiral. For example, on a $3\times3\times6$ cell, stacking the $P3_121$ arrangement with its mirror image gives centrosymmetric arrangements with space groups $P\bar{1}$ and $C2/c$ (Supplementary Information). However, because every chain repeats every three sites, any $3\times3\times3$ region of an arrangement that obeys both rules is one of the twelve arrangements enumerated above, and so is chiral. An arrangement made of chiral regions can be achiral only if it also contains their mirror images, so any achiral arrangement must contain regions of both hands that meet at twin boundaries.

\section{Demonstration in \ce{NbO2F}: DFT and cluster expansion}
\label{sec:dft}

For the chiral phase to be the equilibrium state of a real material, three conditions must hold: first, that the cis and period-three ordering rules are energetically favoured, so that every $3\times3\times3$ region of the $T=0$ ground state is chiral; second, that twin boundaries between regions of opposite hand cost energy, so that the ground state as a whole is homochiral; and third, that this chirality is thermodynamically robust at experimentally relevant temperatures. We show that all three conditions hold for \ce{NbO2F}---the first two here, the third in Section~\ref{sec:finiteT}.

\begin{figure}[tb]
  \centering
  \resizebox{8.5cm}{!}{\includegraphics*{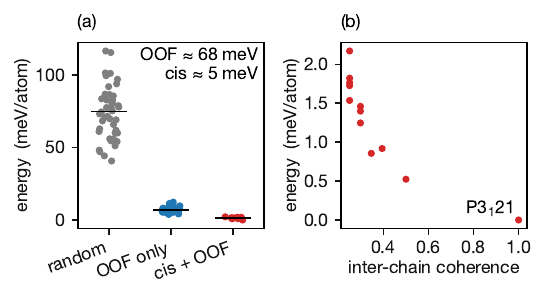}}%
    \caption{\label{fig:dft}
    Density-functional energies of $3\times3\times3$ \ce{NbO2F} configurations, relative to the ground state. (a)~Energy by ensemble: random F/O placements, period-three OOF orderings with random inter-chain phase, and the cis + OOF orbits; bars mark the median energy for each ensemble. (b)~Energy versus inter-chain coherence for all twelve chiral orbits.
    }
\end{figure}

To examine whether the two rules are energetically active in \ce{NbO2F}, we computed density-functional theory energies for $3\times3\times3$ anion configurations (Methods) from three anion ordering classes: random F/O placements, period-three OOF orderings with random inter-chain phases, and the twelve cis + OOF orbit representatives obtained by enumeration (Section~\ref{sec:theorem}). These DFT energies fall into three tiers according to the class of anion ordering (Fig.~\ref{fig:dft}a). Structures with random anion configurations lie a median of $\approx$\qty{75}{\milli\electronvolt\per\atom} above the lowest energy structure, ranging from \qtyrange{41}{117}{\milli\electronvolt\per\atom}. Structures with period-three OOF ordering are significantly more stable, and sit a median of $\approx$\qty{7}{\milli\electronvolt\per\atom} above the lowest-energy structure. The cis + OOF orbit representatives are even more stable: all are within \qty{2.2}{\milli\electronvolt\per\atom} of the lowest-energy structure. OOF ordered structures thus lie lower in energy than disordered ones, and the lowest-energy structures are those exhibiting both OOF and cis orderings. Both rules, therefore, are energetically active in \ce{NbO2F}.

We next compare the twelve chiral orbits by their DFT energies and inter-chain coherence: the degree to which the period-three pattern is correlated between chains (Methods). The energy falls systematically as the coherence increases (Fig.~\ref{fig:dft}b), and the lowest-energy orbit is the maximally coherent orbit, $P3_121$. In this arrangement, the fluoride-free axes cycle $x\to y\to z$ along $\langle111\rangle$, tracing the $3_1$ screw axis, and in the structure itself the Nb--F--Nb bonds wind in helices about this axis (Fig.~\ref{fig:p3121}b,c).

The correlation between coherence and energy can be understood as electrostatic in origin. Each fluoride carries an excess formal charge of $+1$ relative to the oxide framework, so fluorides repel one another. In all twelve orbits, cis-\ce{NbO4F2} coordination means that every cation has one nearest-neighbour F--F pair. The first anion pairs at which the orbits can differ are therefore the next-nearest pairs that sit side by side on adjacent parallel chains. Comparing the DFT energies of the twelve orbits with their numbers of side-by-side F--F pairs gives a linear relationship ($R^2 = 0.97$; Supplementary Information). Orbits that are more coherent generally have fewer side-by-side pairs, so the energy falls as the coherence increases. The $P3_121$ arrangement and its mirror image are the only arrangements with no side-by-side fluoride pairs (Supplementary Information), and are therefore lowest in energy.

Every achiral arrangement that obeys both rules contains twin boundaries between regions of opposite hand (Section~\ref{sec:theorem}). To test whether these boundaries cost energy, we calculated DFT energies for every achiral arrangement on the $3\times3\times6$ cell that obeys both rules (Supplementary Information). All of them lie \qtyrange{0.50}{2.03}{\milli\electronvolt\per\atom} above $P3_121$ in the same cell. Twin boundaries therefore cost energy in \ce{NbO2F}.

To search for lower-energy configurations outside those calculated with DFT, we trained a cluster expansion on the DFT energies of 96 configurations in $6\times6\times6$ supercells (Methods) and performed parallel-tempering Monte Carlo quenches of the resulting model. Independent quenches from random starting configurations all relax to the same $P3_121$ state. The homochiral $P3_121$ arrangement is therefore the predicted ground state of \ce{NbO2F} on three counts: it is the lowest of the arrangements calculated with DFT on the $3\times3\times3$ cell, every achiral arrangement calculated on the $3\times3\times6$ cell lies higher in energy, and the cluster expansion finds no arrangement lower in energy.

\section{Finite-temperature transition}
\label{sec:finiteT}

\begin{figure*}[tb]
  \centering
  \resizebox{\textwidth}{!}{\includegraphics*{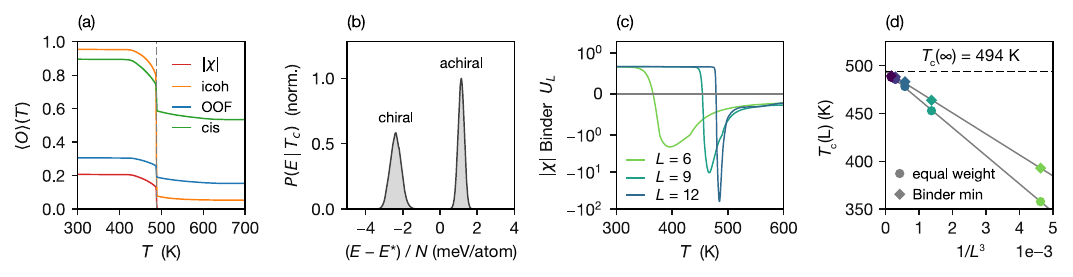}}%
    \caption{\label{fig:finiteT}
    Finite-temperature behaviour of the chiral-to-achiral transition in \ce{NbO2F} from replica-exchange Wang--Landau sampling. (a)~Reweighted order parameters ($|\chi|$, inter-chain coherence, OOF amplitude, and cis fraction) versus temperature at $L=18$. The dashed line marks $T_\mathrm{c}(L=18)$. (b)~Energy distribution $P(E\,|\,T_\mathrm{c})$ for $L=18$ at the equal-weight coexistence temperature. Energies are plotted per atom, relative to the saddle point $E^*$ between the two peaks. (c)~Binder cumulant of the chirality order parameter $\left|\chi\right|$, $U_L(T)$, for $L=6,9,12$. The minima at $L=15$ and $18$ are much deeper and are omitted so that the three smaller sizes remain visible. (d)~Finite-size scaling of the transition temperature, $T_\mathrm{c}(L)$ versus $1/L^3$, for two estimators: the equal-weight coexistence point (circles) and the Binder-cumulant minimum (diamonds), each with a $1/L^3$ fit extrapolating to the common $T_\mathrm{c}$ (dashed line). The uncertainties on $T_\mathrm{c}(L)$ are smaller than the symbols.
    }
\end{figure*}

To assess whether the chirality survives at experimentally relevant temperatures, we performed replica-exchange Wang--Landau Monte Carlo sampling of the cluster-expansion model for $L\times L\times L$ supercells with sizes $L = 6$, $9$, $12$, $15$ and $18$ (Methods), tracking four order parameters across temperature: the cis fraction and OOF amplitude, which measure the strength of the two local orderings; the inter-chain coherence, which measures whether the period-three patterns on different chains hold a fixed phase relationship; and the chirality $\chi$, which measures the net helicity of the anion ordering (Supplementary Information).

Figure~\ref{fig:finiteT}a shows all four order parameters for $L=18$. All change abruptly at the same temperature, suggesting a single first-order phase transition. In particular, $\left|\chi\right|$ is large and roughly constant at low temperature and zero above the transition. The chiral phase is therefore the equilibrium state up to a definite transition temperature, $T_\mathrm{c}$.

To clarify the order of this transition, we first examine the distribution of the system's energy, $P(E\,|\,T)$. Scanning across temperature, we find a narrow range in which this distribution is bimodal, with one peak at the energy of each coexisting phase (Fig.~\ref{fig:finiteT}b, $L=18$). At a first-order transition this coexistence sharpens with system size: the trough between the peaks deepens, because configurations intermediate between the two phases contain an interface whose free-energy cost grows with $L$; and the separation between the peaks, which gives the latent heat, approaches a fixed value per atom. We observe both behaviours (Supplementary Information). The Binder cumulant of the chirality, $U_L$, provides a second, standard test (Methods). At a first-order transition it dips below zero near $T_\mathrm{c}$, with a minimum that deepens with system size, as we observe (Fig.~\ref{fig:finiteT}c). The transition is first order.

The energy distribution and Binder cumulant also provide estimates of the transition temperature at each system size. $T_\mathrm{c}(L)$ can be defined as the temperature at which the two peaks of $P(E\,|\,T)$ carry equal weight, or as the temperature at which $U_L$ reaches its minimum. The two definitions give estimates that differ at each $L$, but each follows the expected first-order finite-size form, $T_\mathrm{c}(L) = T_\mathrm{c} + a/L^{3}$.
Both sets of estimates extrapolate to a common $T_\mathrm{c} = \qty{494.0 \pm 0.7}{\kelvin}$ (Fig.~\ref{fig:finiteT}d).

The behaviour of the four order parameters across $T_\mathrm{c}$ provides additional information about the character of the transition. The cis fraction and OOF amplitude show relatively small proportional changes (Fig.~\ref{fig:finiteT}a), and remain well above their random-disorder reference values above $T_\mathrm{c}$ (Supplementary Information). Both local orderings soften rather than vanish above the transition. The chirality and the inter-chain coherence, by contrast, fall towards zero. What orders at $T_\mathrm{c}$ is therefore the relative phase of the chains. Above the transition the period-three patterns on different chains have no fixed phase relationship, and below it they lock into the relative phase of the helical ground state. The chirality falls with the coherence because $\chi$ can be nonzero only when the chains hold a fixed relative phase. Symmetry analysis shows that the chirality is an improper order parameter, induced by the inter-chain ordering (Supplementary Information).

\section{Generality beyond \ce{NbO2F}}
\label{sec:extension}

The combinatorial theorem of Section~\ref{sec:theorem} assumes only the geometry of the \ce{ReO3}-type network, the 2:1 stoichiometry and the two ordering rules, not the specific cation or anion chemistry. Any \ce{ReO3}-type \ce{MX2Y} system has the same twelve chiral arrangements available. The open question for each material is whether both orderings are energetically favoured.

\ce{TaO2F} is a structural and stoichiometric analogue of \ce{NbO2F}~\cite{FrevelAndRinn_ActaCrystallogr1956}, sharing its \ce{ReO3}-type framework, its 2:1 anion ratio and its $d^0$ cation chemistry. To test whether both orderings are favoured in \ce{TaO2F}, we calculated DFT energies for the twelve chiral cis + OOF arrangements, for eight arrangements with OOF order but random inter-chain phases, and for eight fully random anion arrangements. We also calculated DFT energies for every achiral arrangement on the $3\times3\times6$ cell that obeys both rules (Supplementary Information). As in \ce{NbO2F}, the cis + OOF arrangements lie below all the other computed arrangements, $P3_121$ is the lowest of the twelve, and every achiral arrangement on the $3\times3\times6$ cell lies above $P3_121$. \ce{TaO2F} is therefore also predicted to have a homochiral $P3_121$ ground state.

Other mixed-anion materials share the anion network and the 2:1 ratio of \ce{NbO2F}. In \ce{TiOF2}, oxides take the place of the fluorides of \ce{NbO2F} as the minority anion. All-cis and period-three chain-ordered configurations are among its low-energy computed structures, so both rules appear to be active~\cite{LegeinEtAl_JAmChemSoc2024}. The mechanism identified in this work, in which the two orderings together force chirality, motivates examining \ce{TiOF2} directly. The mixed-anion perovskites add an A-site cation to the same framework. In the oxynitride perovskites, cis coordination is well established~\cite{YangEtAl_NatChem2011,PageEtAl_ChemMater2007,Attfield_CrystGrowthDes2013}, while evidence for period-three O--O--N ordering is currently lacking. Period-three ordering requires equal amounts of nitride on the sites along all three axes. In \ce{SrTaO2N}, \ce{SrNbO2N} and related oxynitrides, diffraction data show that almost all the nitride occupies anion sites along only two of the three axes~\cite{YangEtAl_NatChem2011,OroSoleEtAl_ChemCommun2013,JohnstonEtAl_ChemCommun2018,Attfield_CrystGrowthDes2013}, which rules out period-three ordering in these materials. In \ce{BaTaO2N} and other cubic oxynitrides, by contrast, the nitride is shared equally between the sites along the three axes on average~\cite{PageEtAl_ChemMater2007,JohnstonEtAl_ChemCommun2018}. This average distribution is consistent with period-three ordering, but whether the nitride is ordered with period three cannot be determined from average-structure data.

The same twelve chiral arrangements can also arise when vacancies take the role of the minority species. In \ce{BaZnO2}~\cite{vonSchneringEtAl_ZAnorgAllgChem1960} and \ce{BaCoO2}~\cite{Spitsbergen_ActaCrystallogr1960,DiattaEtAl_JAmChemSoc2024}, oxygen and vacancies share the anion sites of a perovskite in a 2:1 ratio. Mapping the published structures of both compounds onto an ideal perovskite shows that the oxygen and vacancies adopt the $P3_121$ arrangement predicted for \ce{NbO2F} (Supplementary Information). Another example is the antiperovskite $\beta$-\ce{Fe2SeO}, in which iron atoms and vacancies occupy the same network of sites in a 2:1 ratio~\cite{ValldorEtAl_AngewChemIntEd2016}. Mapping the published structure~\cite{ShamovaEtAl_ChemMater2026} onto an ideal antiperovskite shows that iron atoms and vacancies adopt the $P3_1$ arrangement (Supplementary Information). In each material, the arrangement of occupied and vacant sites maps onto one of the twelve chiral arrangements allowed by combined cis coordination and period-three chain order (Section~\ref{sec:theorem}). The chiral structures of \ce{BaZnO2}, \ce{BaCoO2} and $\beta$-\ce{Fe2SeO} can therefore be understood as a consequence of these two short-range ordering rules.

\section{Experimental realisation and detection}
\label{sec:consequences}

The predictions of Sections~\ref{sec:dft} and \ref{sec:finiteT} await experimental test: confirming them requires synthesising the chiral phase and then distinguishing it from the achiral phase. Synthesis below $T_\mathrm{c}$ is expected to yield the equilibrium chiral phase directly. Synthesis above $T_\mathrm{c}$ would instead produce the achiral phase, which would then require annealing below $T_\mathrm{c}$ so that the anions can reorganise into the chiral arrangement. Whether either route is practical depends on the kinetics of anion ordering below $T_\mathrm{c}$.

Once formed, the chiral and achiral phases can be distinguished by diffraction. Both phases produce $\{hk1/3\}^{*}$ superlattice features from their period-three chain ordering, but differ in the sharpness of these features. In the achiral phase the chains are mutually uncorrelated, so the features are sheets: narrow along each chain direction, diffuse transverse to it; in the chiral phase the chains have a fixed phase relationship, and the sheets sharpen towards spots~\cite{KeenAndGoodwin_Nature2015}. The sharpness of these features thus directly measures the inter-chain coherence (Fig.~\ref{fig:finiteT}a). The two cases differ in the shape of these features, not their positions, so discriminating between the two phases requires single-crystal or electron diffraction. Previous electron diffraction data for \ce{NbO2F} show diffuse sheets~\cite{BrinkEtAl_JSolidStateChem2002}, indicating that those samples were likely in the achiral phase.

\section{Conclusions}
\label{sec:conclusions}
We have shown that structural chirality can be forced by the short-range order of two anion species over the sites of an achiral parent structure, independent of atomic displacements. In an achiral \ce{ReO3}-type parent with two anions in a 2:1 ratio, two short-range rules, cis coordination and period-three chain ordering, together allow only twelve symmetry-inequivalent arrangements on the smallest commensurate cell, and all twelve are chiral. In any larger structure that obeys both rules, every $3\times3\times3$ region matches one of these twelve arrangements, so the rules force chirality locally at any size. For \ce{NbO2F}, density-functional theory calculations predict that twin boundaries between regions of opposite hand cost energy, so that the ground state is homochiral. This ground state is the $P3_121$ arrangement, which maximises the separation between fluorides on adjacent parallel chains. Replica-exchange Wang--Landau sampling of a DFT-trained cluster-expansion model predicts that the chiral phase is the equilibrium state up to a first-order transition at \qty{494}{\kelvin}. The transition orders the relative phases of the period-three anion chains, and the chirality appears and vanishes with this inter-chain ordering.

The result that the two rules together force chirality depends only on the geometry of the \ce{ReO3}-type network, the 2:1 stoichiometry and the rules themselves, not on any specific chemistry. It therefore extends to any system with this network and stoichiometry in which both orderings are active, including systems in which vacancies take the role of the minority species. We also predict a homochiral $P3_121$ ground state for \ce{TaO2F}, and find that, for three known vacancy-ordered compounds, \ce{BaZnO2}, \ce{BaCoO2} and $\beta$-\ce{Fe2SeO}, each structure maps onto one of the twelve chiral arrangements. The chirality of these compounds can therefore be understood in terms of the two short-range rules. These known structures, together with the predictions for \ce{NbO2F} and \ce{TaO2F}, establish short-range order as a route to structural chirality and mark heteroanionic and vacancy-ordered materials in which both orderings are active as candidates for new chiral phases.

\section{Methods}
\label{sec:methods}

\subsection{Density-functional calculations}
Total energies were computed with plane-wave density-functional theory (\textsc{vasp}~\cite{KresseAndFurthmuller_PhysRevB1996,KresseAndJoubert_PhysRevB1999}) using the PBEsol exchange--correlation functional~\cite{PerdewEtAl_PhysRevLett2008}, projector-augmented-wave potentials (\ce{Nb}\_sv, \ce{O}, \ce{F}), a \qty{600}{\electronvolt} plane-wave cutoff, Gaussian smearing ($\sigma = \qty{0.05}{\electronvolt}$) and an electronic-energy tolerance of \qty{e-6}{\electronvolt}, with full relaxation of the cell and ionic positions to a residual force below \qty{0.01}{\electronvolt\per\angstrom}. Calculations on the $3\times3\times3$ cell (108 atoms) used a $2\times2\times2$ Monkhorst--Pack $k$-point mesh. Calculations on the $3\times3\times6$ cell (216 atoms) used a $2\times2\times1$ $k$-point mesh offset by $(\tfrac12,\tfrac12,\tfrac12)$ from $\Gamma$. The cluster-expansion training set used $6\times6\times6$ supercells with $\Gamma$-point sampling. To test whether the energy ordering of the twelve orbits changes under a second exchange--correlation functional, we performed additional relaxations of all twelve using r$^2$SCAN~\cite{FurnessEtAl_JPhysChemLett2020}, starting from their PBEsol structures. The orbits fall in the same order as with PBEsol, with energy splittings 0.79--0.88 times the PBEsol values.
\subsection{Cluster expansion}
A cluster expansion of the energy as a function of the anion arrangement was constructed with \textsc{icet}~\cite{AngqvistEtAl_AdvTheorySimul2019}. The fitting method (automatic relevance determination regression, ARDR), the pair, triplet and quadruplet cutoffs (\qtylist{9;5;5}{\angstrom}) and the ARDR relevance threshold were selected by cross-validated RMSE with \textsc{trainstation}~\cite{FranssonEtAl_npjComputMater2020}. The effective cluster interactions were then fitted to the PBEsol energies on standardised inputs, using scikit-learn~\cite{PedregosaEtAl_JMachLearnRes2011}. The training set was assembled iteratively. Early rounds targeted the low-energy orderings: each round, parallel tempering on the expansion fitted from the data gathered so far identified its lowest-energy configurations, which were computed with DFT and added. This continued until repeated parallel-tempering runs returned the same lowest-energy configuration. Later rounds used uncertainty-driven active learning~\cite{KleivenEtAl_JPhysEnergy2021}: ARDR returns a Gaussian posterior over the effective cluster interactions rather than a single best-fit value for each, and drawing sets of interactions from this posterior gives a committee of expansions whose energy predictions coincide where the training data are informative and diverge where they are not. Each expansion in the committee predicted the energies of configurations visited in parallel-tempering runs. The structures on which they disagreed most were computed with DFT and added, subject to a minimum-Hamming-distance filter on the anion occupations that excluded near-duplicates. The final training set comprised 96 configurations in $6\times6\times6$ supercells, and the model has a cross-validated RMSE of \qty{1.4}{\milli\electronvolt\per\atom}. We also tested the expansion against the DFT energies of the eleven chiral orbits other than $P3_121$, none of which is in the training set. These orbits lie within \qty{2.2}{\milli\electronvolt\per\atom} of $P3_121$. For these orbits, the root-mean-square error of the expansion is \qty{0.14}{\milli\electronvolt\per\atom}. The largest error, \qty{0.32}{\milli\electronvolt\per\atom}, is smaller than the \qty{0.52}{\milli\electronvolt\per\atom} gap between $P3_121$ and the nearest other orbit.

\subsection{Ground-state search}
Candidate ground states were located by canonical Monte Carlo on $6\times6\times6$ supercells (864 atoms), using \textsc{mchammer}~\cite{AngqvistEtAl_AdvTheorySimul2019} with custom trial moves for the anion sublattice, combined with parallel-tempering exchange across a temperature ladder. Six move types were used: pair swaps (exchange any two anions), row shifts (slide the anion pattern of a $\langle001\rangle$ chain by one site), motif shifts (cycle a single three-anion period within a chain by one site), chain swaps (exchange the anion patterns of two parallel chains), row reflections (reflect a chain's anion pattern along its length) and cell reflections (reflect the full cell in a randomly chosen $\langle001\rangle$ mirror plane, mapping a configuration to its mirror image). The collective moves allow the chain ordering to reorganise at temperatures where pair swaps alone become kinetically trapped. Independent quenches from random starting configurations converged to the same chiral ground state. To check that the expansion reproduces the DFT ground state on cells of any size, we split its energy into contributions from $3\times3\times3$ windows (Supplementary Information). Among arrangements that obey both rules, its lowest-energy arrangement on a cell of any size is homochiral $P3_121$.

\subsection{Replica-exchange Wang--Landau simulations}
Finite-temperature behaviour was characterised by replica-exchange Wang--Landau (REWL) sampling~\cite{WangAndLandau_PhysRevLett2001,WangAndLandau_PhysRevE2001,VogelEtAl_PhysRevLett2013}, implemented in \textsc{mchammer-pt}~\cite{Morgan_mchammerpt} on top of \textsc{mchammer}'s Wang--Landau ensemble and the same custom trial moves. Simulation cells were $L\times L\times L$ supercells of the cubic unit cell, containing $4L^3$ atoms. For each system size $L=6$, $9$, $12$, $15$ and $18$, the energy range spanning the coexistence region was partitioned into overlapping windows, each sampled by independent Wang--Landau walkers whose density-of-states estimates were merged. Each walker followed the standard flatness-controlled schedule for the modification factor, crossing over continuously to the $1/t$ schedule of Belardinelli and Pereyra~\cite{BelardinelliAndPereyra_PhysRevE2007,BelardinelliAndPereyra_JChemPhys2007}, once the flatness-driven updates reached the $1/t$ curve, with $t$ counted as the number of trial moves since the walker reached its energy window. Implementation details are provided in \textsc{mchammer-pt}~\cite{Morgan_mchammerpt} and the deposited run inputs. Per-window densities of states were stitched on their overlaps into a single $\ln g(E)$ per run, and five independent seeds were run at each size. Runs were extended until the equal-weight coexistence temperature (below) was stable under further sampling, consistently across the five seeds.

The coexistence temperature was obtained from each stitched $\ln g(E)$ by the equal-weight construction, locating the temperature at which the two phases carry equal canonical weight about the saddle of $\phi(E)=\beta E-\ln g(E)$. To locate the two peaks and the saddle without locking onto bin-scale noise in the stitched densities of states, $\ln g(E)$ was smoothed with a Gaussian a few energy bins wide. The equal-weight sums themselves use the unsmoothed $\ln g(E)$. For each run, the smoothing width was the smallest of 2, 4, 8, 16 and 30 bins whose $T_\mathrm{c}$ agrees to within \qty{1}{\kelvin} with that at the next larger width, among widths that give a latent heat within 10\% of the largest found. The selected widths are 2 bins for every run at $L=9$ and $12$, 2--8 bins at $L=15$ and 4--8 bins at $L=18$, against peak separations of about 455 and 822 bins at $L=15$ and $18$ (bin width \qty{0.1}{\electronvolt}). At $L=6$, where the peaks are only 29 bins apart (bin width \qty{0.05}{\electronvolt}), only the 2-bin width keeps them resolved, so the $L=6$ estimate is not checked against a larger width. Removing $L=6$ from the extrapolation moves $T_\mathrm{c}$ by \qty{0.15}{\kelvin}. An independent per-size estimate of the transition temperature was taken from the minimum of the Binder cumulant of the chirality order parameter, $U_L = 1-\langle \chi^{4}\rangle/(3\langle \chi^{2}\rangle^{2})$~\cite{Binder_ZPhysB1981,BinderAndLandau_PhysRevB1984}. 

The equal-weight and Binder cumulant $T_\mathrm{c}$ estimates were separately extrapolated to the thermodynamic limit using the first-order finite-size form $T_\mathrm{c}(L)=T_\mathrm{c}+a/L^{3}$. Fitting was by Bayesian regression. The modelled uncertainty in each $T_\mathrm{c}(L)$ has two contributions: the seed-to-seed standard error ($\lesssim$\qty{0.2}{\kelvin}), and a scatter $\sigma_{\mathrm{sys}}$ that allows individual sizes to deviate from the $1/L^{3}$ form. The priors for $T_\mathrm{c}$ and $a$ were flat, and the prior for $\sigma_{\mathrm{sys}}$ was half-normal with scale \qty{1}{\kelvin}. The fit returns $\sigma_{\mathrm{sys}} = \qty{1.1}{\kelvin}$. We report the posterior mean and standard deviation, $T_\mathrm{c} = \qty{494.0 \pm 0.7}{\kelvin}$. The two estimators converge to a common limit, and the extrapolated $T_\mathrm{c}$ is stable, within its quoted uncertainty, against removing any single system size from the fit. The largest shift, on removing $L=18$, is \qty{0.5}{\kelvin}.

\subsection{Structural order parameters}
Canonical averages of structural order parameters were recovered from the converged densities of states by a frozen-$g$ measurement pass: holding $\ln g(E)$ fixed, the walkers sampled each window, accumulating the microcanonical moments $\langle O\rangle(E)$ in each energy bin. These moments were reweighted to canonical averages, $\langle O\rangle(T)=\sum_E \langle O\rangle(E)\,g(E)\,\mathrm{e}^{-E/k_\mathrm{B}T}/Z$, from which the Binder cumulants were computed. Four structural order parameters were measured. The OOF amplitude is the chain-averaged magnitude of the period-three Fourier component of the binary fluoride occupation along the $\langle001\rangle$ chains, normalised by chain length so that ideal period-three ordering gives $1/3$. The inter-chain coherence is the amplitude-weighted spatial autocorrelation of that per-chain Fourier coefficient across the chain plane, averaged over non-zero lags and the three chain directions. It equals unity when every chain holds a fixed phase relative to its neighbours, as in the helical $x\to y\to z$ stacking of $P3_121$, and tends to zero for uncorrelated phases in the large-cell limit. Both are computed from the per-chain Fourier coefficients provided by the \textsc{chainorder} library~\cite{Morgan_chainorder}. The cis fraction is the fraction of cations coordinated by two fluorides in a cis arrangement (cis-\ce{NbO4F2}). The chirality order parameter $\chi$ is the pseudoscalar invariant $|E_+|^2-|E_-|^2$ of the $\langle111\rangle$ anion ordering (Supplementary Information). Its sign is the handedness and its magnitude is the degree of chiral order. It is evaluated as the projection onto the pseudoscalar representation of the cubic point group, which makes it a proper pseudoscalar for any configuration, invariant under proper rotations and sign-reversing under reflection. Since $\langle\chi\rangle$ vanishes by symmetry, canonical averages are reported as the root-mean-square $\sqrt{\langle\chi^2\rangle}$ (written $|\chi|$ in the figures). The cis fraction was computed directly from the anion arrangement with \textsc{nbo2f\_analysis}, and $\chi$ with \textsc{chainorder}. Reference-state values for these order parameters, with their random-disorder limits, are provided in the Supplementary Information.

\emph{Code and data availability.} The replica-exchange Wang--Landau orchestrator and analysis CLIs (\textsc{mchammer-pt}) and the order-parameter library (\textsc{chainorder}) are openly available on PyPI. Both packages, together with the project driver (\textsc{nbo2f\_analysis}), are hosted at \url{https://github.com/bjmorgan}. Figure~\ref{fig:ReO3} and panels a and c of Fig.~\ref{fig:p3121} were drawn with \textsc{hofmann}~\cite{Morgan_hofmann}. All calculation inputs and outputs (DFT and Monte Carlo), the cluster-expansion fitting, and the analysis and figure-generation scripts are deposited at Zenodo (\url{https://doi.org/10.5281/zenodo.21799772}).

\section{Acknowledgements}
This work used the ARCHER2 UK National Supercomputing Service (\url{https://www.archer2.ac.uk})~\cite{BeckettEtAl_Zenodo2024} through our membership of the UK's HEC Materials Chemistry Consortium, funded by EPSRC (EP/X035859).

\bibliography{bibliography}

\ifsubmission\else
  \clearpage
  \section*{Supplementary Information}
  \setcounter{secnumdepth}{2}
\setcounter{section}{0}
\setcounter{figure}{0}
\setcounter{table}{0}
\renewcommand{\thesection}{S\arabic{section}}
\renewcommand{\thefigure}{S\arabic{figure}}
\renewcommand{\thetable}{S\arabic{table}}

\section{Enumeration of cis + OOF configurations}
\label{sec:si-enumeration}

The enumeration of Section~\ref{sec:theorem} is a depth-first search over cation labellings, in which each cis cation is labelled by its fluoride-free axis (Fig.~\ref{fig:p3121}a). The search relies on two properties of this labelling: that OOF ordering requires each chain to contain exactly one cation whose label lies along the chain direction, and that the labellings satisfying this one-per-chain condition correspond one-to-one with the cis + OOF configurations. We demonstrate both here.

Within the $3\times3\times3$ supercell, each $\langle001\rangle$ chain contains three cations and three anions. Under OOF ordering two cations have along-chain (O,F) neighbours, and one cation has along-chain (O,O) neighbours. An OOF-ordered chain therefore contains exactly one cation whose fluoride-free axis---and hence whose label---lies along the chain direction (an own-axis label). Chains with zero own-axis labels, or with two or more, cannot be OOF-ordered (Fig.~\ref{fig:si-rowtest}). With zero own-axis labels, all three cations require at least one fluoride along-chain neighbour, which requires at least two fluorides. With two or more own-axis labels, each labelled cation requires both of its chain anions to be oxide. The chain has only three anions, so two labelled cations force all three to oxide, leaving no site for the chain's fluoride. A labelling is therefore consistent with OOF ordering only if every chain contains exactly one own-axis label. We call such labellings valid.

The search enumerates labellings, but the theorem concerns anion configurations, so the two must correspond exactly. A valid labelling determines its configuration completely: in each chain the fluoride must occupy the anion between the two cations not labelled with the chain direction, and because no anion belongs to two chains, every fluoride is placed exactly once and without conflict. A configuration built this way also matches the labels it was built from. Along each cation's labelled axis, that cation is the chain's unique own-axis-labelled cation, so both of its anions on that axis are oxide; along its other two axes, it is adjacent to a placed fluoride. Each cation therefore has exactly two fluorides, in a cis arrangement, and its fluoride-free axis is its label. In the reverse direction, every cis + OOF configuration gives each cation a single fluoride-free axis, which is its labelling. Valid labellings and cis + OOF configurations therefore correspond one to one.

The search returns \num{10752} valid labellings. Their configurations reduce under the parent symmetry to the twelve orbits of Section~\ref{sec:theorem}, shown as label patterns in Fig.~\ref{fig:si-gallery}.

\begin{figure}[tb]
  \centering
  \resizebox{7.9cm}{!}{\includegraphics*{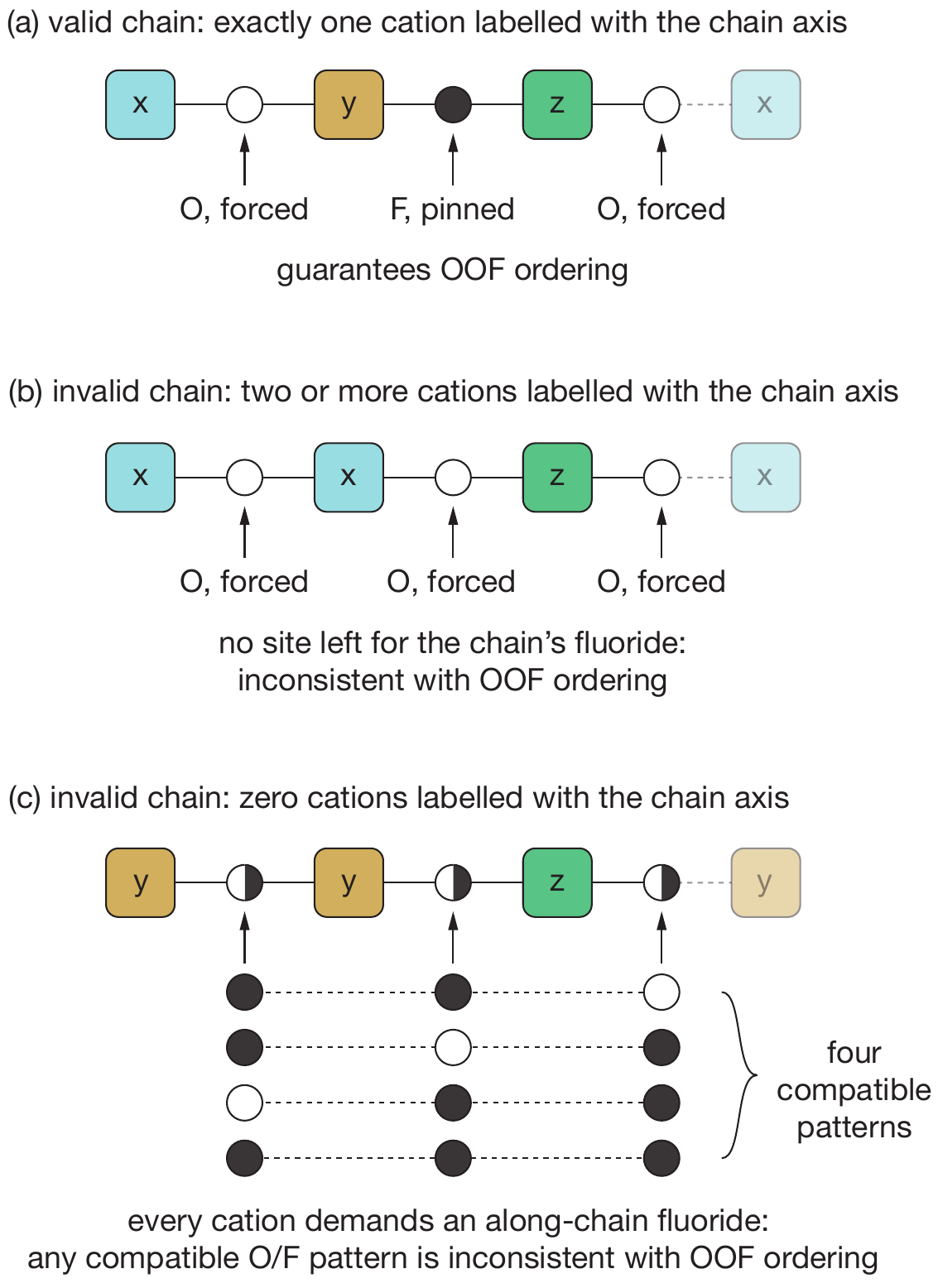}}%
  \caption{\label{fig:si-rowtest} The chain test of the labelling encoding. Each panel shows one $\langle001\rangle$ chain of the supercell (three cations and three shared anions; the faded cation repeats the first, showing the periodic wrap), with cations coloured by fluoride-free-axis label ($x$ blue, $y$ gold, $z$ green, as in Fig.~\ref{fig:p3121}), fluoride black and oxide open. (a)~A valid chain: exactly one cation has an own-axis label, its two chain anions are forced to oxide, and the chain's fluoride is pinned to the remaining site. (b)~Two or more own-axis labels force all three anions to oxide, leaving no site for the chain's fluoride. (c)~With no own-axis label, every cation demands a fluoride along the chain, and each of the four compatible anion patterns contains two or more fluorides. Chains of type (b) or (c) are inconsistent with OOF ordering, and are pruned by the search.}
\end{figure}

\begin{figure*}[p]
  \centering
  \resizebox{!}{0.85\textheight}{\includegraphics*{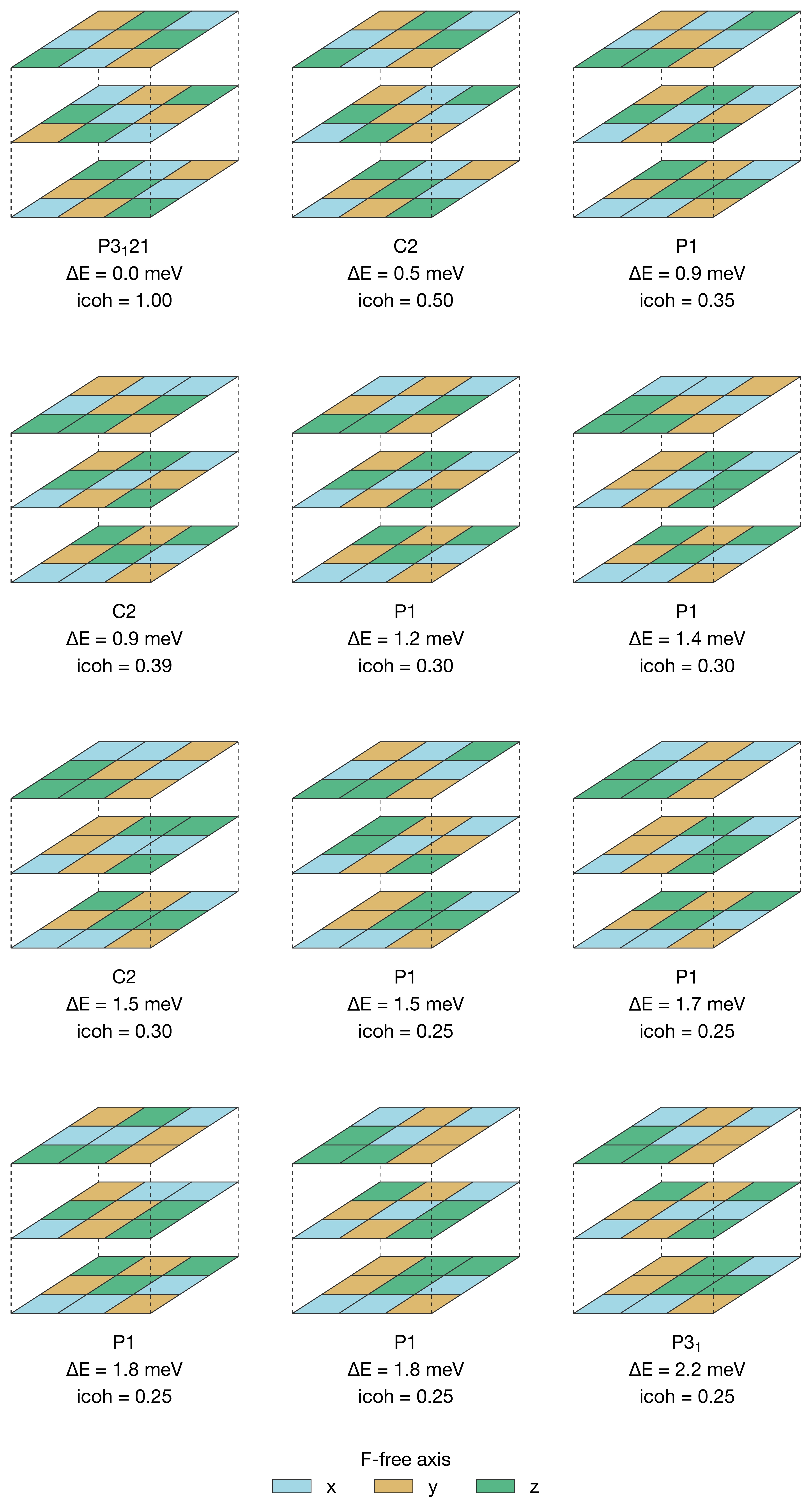}}%
  \caption{\label{fig:si-gallery} The twelve cis\,+\,OOF orbits as fluoride-free-axis label patterns. Each orbit's $3\times3\times3$ cation cell is coloured by fluoride-free axis ($x$ blue, $y$ gold, $z$ green, as in Fig.~\ref{fig:p3121}) and shown as three stacked $\langle001\rangle$ layers, labelled with its Sohncke space group ($P3_121$, $P3_1$, $C2$ or $P1$), its DFT energy above $P3_121$ (per atom), and its inter-chain coherence (icoh). Orbits are ordered by that energy (top-left to bottom-right); all twelve are chiral, with $P3_121$ the maximally symmetric. The coherence reaches unity only for the $P3_121$ helix.}
\end{figure*}

\section{Arrangements on the \texorpdfstring{$3\times3\times6$}{3x3x6} cell}
\label{sec:si-336}

In the main text, we give stackings of $P3_121$ with its mirror image on the $3\times3\times6$ cell as examples of achiral arrangements that obey both rules (Section~\ref{sec:theorem}), and compare the DFT energies of every such achiral arrangement with that of $P3_121$ (Section~\ref{sec:dft}). Here we describe the enumeration of these arrangements and give their DFT energies. In any arrangement that obeys both rules, every $3\times3\times3$ region is one of the twelve orbits enumerated in Section~\ref{sec:theorem}. An arrangement on the $3\times3\times6$ cell that obeys both rules is therefore a stack of (001) planes taken from the valid $3\times3\times3$ labellings. To find all such arrangements, we enumerated every sequence of six of these planes and kept those in which every chain along $z$ has OOF order. These arrangements reduce to 130 symmetry-inequivalent classes, of which fifteen are achiral. Table~\ref{tab:si-336} gives the DFT-calculated energies of all fifteen achiral arrangements for \ce{NbO2F} and \ce{TaO2F}, relative to the homochiral $P3_121$ arrangement in the same cell.

\begin{table}[tb]
  \caption{\label{tab:si-336} The fifteen achiral classes of arrangements that obey both rules on the $3\times3\times6$ cell, in order of their \ce{NbO2F} DFT energy. Stacking: for the three stackings of the $P3_121$ arrangement and its mirror image, 0 marks a plane of $P3_121$ and 1 a plane of its mirror image. $\Delta E$: DFT energy above the homochiral $P3_121$ arrangement in the same cell, for \ce{NbO2F} and \ce{TaO2F}.}
  \begin{ruledtabular}
  \begin{tabular}{llcc}
    & & \multicolumn{2}{c}{$\Delta E$ (meV/atom)} \\
    Stacking & Space group & \ce{NbO2F} & \ce{TaO2F} \\
    \hline
    000111 & $C2/c$ & 0.50 & 0.46 \\
    010101 & $C2/c$ & 0.55 & 0.35 \\
    001011 & $P\bar{1}$ & 0.56 & 0.43 \\
    -- & $P\bar{1}$ & 0.87 & 0.79 \\
    -- & $Pc$ & 1.04 & 0.91 \\
    -- & $P\bar{1}$ & 1.07 & 0.94 \\
    -- & $P\bar{1}$ & 1.15 & 0.99 \\
    -- & $C2/c$ & 1.26 & 1.11 \\
    -- & $P\bar{1}$ & 1.26 & 1.06 \\
    -- & $Cc$ & 1.29 & 1.07 \\
    -- & $P\bar{1}$ & 1.40 & 1.15 \\
    -- & $Pc$ & 1.41 & 1.15 \\
    -- & $C2/c$ & 1.54 & 1.18 \\
    -- & $Cc$ & 1.74 & 1.39 \\
    -- & $Cc$ & 2.03 & 1.59 \\
  \end{tabular}
  \end{ruledtabular}
\end{table}

\section{Side-by-side fluoride pairs}
\label{sec:si-electrostatics}

In the main text, we explain the DFT energies of the twelve cis + OOF arrangements of \ce{NbO2F} in terms of their numbers of side-by-side fluoride pairs (Section~\ref{sec:dft}). Figure~\ref{fig:si-electrostatics} plots these energies against the number of side-by-side pairs in each arrangement.

\begin{figure}[tb]
  \centering
  \resizebox{4.27cm}{!}{\includegraphics*{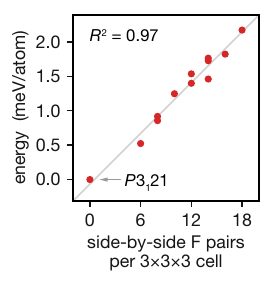}}%
  \caption{\label{fig:si-electrostatics} DFT energies of the twelve cis + OOF orbits of \ce{NbO2F}, relative to $P3_121$, against the number of side-by-side fluoride pairs in the $3\times3\times3$ cell. The dashed line is the least-squares fit.}
\end{figure}

\section{The cluster-expansion energy at any cell size}
\label{sec:si-ce-bound}

In the main text, we show that every $3\times3\times3$ region of an arrangement that obeys both rules is one of the twelve orbits (Section~\ref{sec:theorem}), and that, in \ce{NbO2F}, $P3_121$ is the lowest in energy of these orbits on the $3\times3\times3$ cell (Section~\ref{sec:dft}). On larger cells, different regions of one arrangement can belong to different orbits, or to both hands. Here we show that, in the cluster expansion used for the Monte Carlo simulations, the lowest-energy arrangement that obeys both rules is $P3_121$ of a single hand, on a cell of any size or shape.

To show this, we assign each cation a window: the $3\times3\times3$ block of cations centred on it, together with the 108 anions bonded to them. We then split the energy of an arrangement into a sum of window energies, each of which depends only on the arrangement inside its window. The cluster expansion writes the energy as a sum over clusters of up to four anion sites, no more than \qty{9}{\angstrom} across (Methods), and every cluster fits inside a window. We shared the energy of each cluster equally among the windows that contain it, so the window energies sum exactly to the total energy. In an arrangement that obeys both rules, every window is a $3\times3\times3$ block of one of the twelve orbits, so only \num{10752} windows are possible, one for each valid labelling (Section~\ref{sec:theorem}). We computed the energy of each.

Every $P3_121$ window has the same energy, and every other window is higher in energy by at least \qty{0.31}{\milli\electronvolt\per\atom}. The energy of an arrangement that obeys both rules is therefore at least that of $P3_121$, and equals it only when every window is a $P3_121$ window. Two neighbouring $P3_121$ windows always belong to the same helix, so an arrangement in which every window is a $P3_121$ window is $P3_121$ of a single hand throughout.

\section{Symmetry of the chiral ordering}
\label{sec:si-symmetry}

The chiral ground state is characterised by a helical arrangement of Nb--F--Nb bonds (Fig.~\ref{fig:p3121}), which cycles $x\to y\to z$ along the body diagonal (or $z\to y\to x$ in the enantiomer) with wavevector $\mathbf{k}=(\tfrac13,\tfrac13,\tfrac13)$. This helical arrangement is formed from the fluoride occupations on the three anion sublattices (the $x$-, $y$- and $z$-oriented bonds), with the occupation on each sublattice ordering at the same wavevector $\mathbf{k}$. The three sublattice orderings are related by the helix's three-fold screw axis, so they have the same amplitude, and their phases advance by a third of a turn from one sublattice to the next. The direction of this phase advance, $x\to y\to z$ or $z\to y\to x$, is the handedness of the helix.

Any anion ordering can be projected onto two perfect helices, one of each handedness. Each sublattice's fluoride occupation has a Fourier component at $\mathbf{k}$ with complex amplitude $g_x$, $g_y$, or $g_z$. Projecting the occupations onto the two helices gives amplitudes $E_+ = g_x + \omega g_y + \omega^2 g_z$ and $E_- = g_x + \omega^2 g_y + \omega g_z$ ($\omega=e^{2\pi i/3}$). Their total, $|E_+|^2+|E_-|^2$, is the helical amplitude, measuring the degree of helical order regardless of handedness. This amplitude is largest in the ground state and zero when disordered. Their difference, $|E_+|^2-|E_-|^2$, is the helical chirality, the order parameter $\chi$ of the main text. A reflection swaps $E_+$ and $E_-$, so their difference changes sign while their total does not. The helical chirality is therefore a pseudoscalar, whose magnitude gives the degree of chiral order and whose sign distinguishes the two enantiomers. The chirality can be non-zero only when the helical amplitude is, so it vanishes wherever the helical order does.

The symmetry at $\mathbf{k}$ also determines whether the chirality is a primary order parameter, driving the transition itself, or an improper one, induced by the anion ordering. The little group of $\mathbf{k}$ (the cubic operations that leave $\mathbf{k}=(\tfrac13,\tfrac13,\tfrac13)$ unchanged) is $C_{3v}$: the three-fold rotation about $\langle111\rangle$, together with three mirrors. Any chirality measure must be a pseudoscalar of this group, transforming as its $A_2$ representation. If an $A_2$ component can be constructed as a linear combination of the ordering amplitudes (first order), the chirality is a primary order parameter; if an $A_2$ component appears only in products of two amplitudes (second order), the chirality is an improper order parameter. Under $C_{3v}$ the three sublattice amplitudes decompose as $A_1\oplus E$: the $A_1$ component is the uniform combination $g_x+g_y+g_z$, and the two $E$ components are $E_+$ and $E_-$, defined above. Because $A_1\oplus E$ excludes $A_2$, no pseudoscalar can be formed at first order, and the chirality cannot be a primary order parameter. The quadratic combinations of $E_+$ and $E_-$ span $E\otimes E = A_1\oplus A_2\oplus E$, which does contain an $A_2$ component: this is the chirality $\chi = |E_+|^2-|E_-|^2$, while the $A_1$ component is the helical amplitude $|E_+|^2+|E_-|^2$. The chirality is therefore an improper order parameter, appearing only when the chains order, as observed in our simulations (Section~\ref{sec:finiteT}).

Section~\ref{sec:finiteT} reports the inter-chain coherence rather than the helical amplitude. The helical amplitude is large only when the chains lock specifically into a helix, of either hand. The inter-chain coherence measures something more general: whether the period-three ordering on each chain holds a fixed phase relative to its neighbours. It reaches unity for any fixed inter-chain phase relationship, helical or not. The helix is one such arrangement, and so is fully coherent. But a structure can be fully coherent without being a helix. Chains with equal phase would be just as coherent, yet carry no helical amplitude. Among the cis + OOF configurations, however, the helix is the only fully coherent state: the inter-chain coherence of the twelve orbit representatives reaches unity only for $P3_121$, with no other orbit above $0.50$ (Fig.~\ref{fig:si-gallery}). The ordered state the simulations find is the helix (Section~\ref{sec:finiteT}), so the helical amplitude and the inter-chain coherence rise and fall together. The inter-chain coherence is the more readily interpreted of the two, so it is the measure chosen for Fig.~\ref{fig:finiteT}.

\section{Size dependence of the coexistence energy distribution}
\label{sec:si-pofe}

Section~\ref{sec:finiteT} identifies two size-dependent signatures of a first-order transition: a deepening trough between the coexistence peaks, and a latent heat that approaches a fixed value per atom. Both strengthen with increasing system size. The energy distribution at coexistence is bimodal at every size we simulated (Fig.~\ref{fig:si-pe-size}a). The trough between the two peaks deepens rapidly with $L$: measured as $\ln[P_\mathrm{peak}/P_\mathrm{trough}]$, its depth is $0.28$, $1.41$, $4.37$, $9.60$ and $17.75$ for $L = 6$, $9$, $12$, $15$ and $18$. At $L=6$ the two phases are barely resolved, and the probability falls by only a quarter between the peaks; at $L=18$ it falls by more than seven orders of magnitude. The separation of the two peaks in the energy distribution is the latent heat per atom (Fig.~\ref{fig:si-pe-size}b). This latent heat rises through \qtylist{1.67;2.52;3.09;3.37;3.52}{\milli\electronvolt\per\atom} in increments that shrink at each step, so it is approaching a bulk value from below.

\begin{figure}[tb]
  \centering
  \resizebox{8.8cm}{!}{\includegraphics*{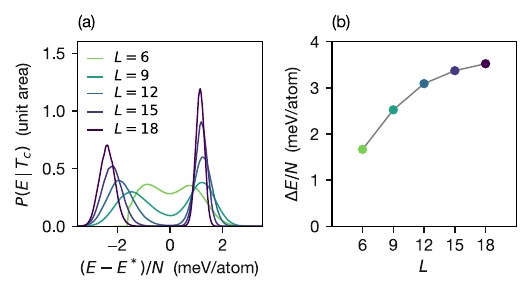}}%
  \caption{\label{fig:si-pe-size} Size dependence of the coexistence energy distribution. (a)~$P(E\,|\,T_\mathrm{c}(L))$ for $L = 6$, $9$, $12$, $15$ and $18$, each computed from the density of states at that size averaged over the five production runs, at that size's equal-weight temperature; energies are plotted per atom, relative to the saddle point $E^*(L)$ between the two peaks. (b)~Latent heat per atom, $\Delta E/N$, from the separation of the two peaks in (a).}
\end{figure}

\section{Structural order-parameter reference values}
\label{sec:op-refs}

The order parameters of Fig.~\ref{fig:finiteT}a are bounded by two reference states: the $P3_121$ ground state and the random-disorder limit. Table~\ref{tab:op-refs} gives the value of each order parameter in these two limits, together with its random-disorder value at $L=18$, the size plotted in Fig.~\ref{fig:finiteT}a.

\begin{table}[tb]
\caption{\label{tab:op-refs} Reference values of the four structural order parameters of Fig.~\ref{fig:finiteT}a at fluoride fraction $f=1/3$: the $P3_121$ ground state, and the random-disorder values in the $L\to\infty$ limit and at $L=18$ from Monte-Carlo sampling.}
\renewcommand{\arraystretch}{1.15}
\begin{tabular}{lccc}
\toprule
order parameter & $P3_121$ & \multicolumn{2}{c}{random} \\
\cmidrule(lr){3-4}
 & & $L\to\infty$ & $L=18$ \\
\midrule
chirality $|\chi|$ & $1/4$ & $0$ & $7.8\times10^{-5}$ \\
inter-chain coherence & $1$ & $0$ & $0.049$ \\
OOF amplitude & $1/3$ & $0$ & $0.098$ \\
cis fraction & $1$ & $64/243$ & $0.263$ \\
\bottomrule
\end{tabular}
\end{table}

\section{DFT energies of \texorpdfstring{\ce{TaO2F}}{TaO2F}}
\label{sec:si-tao2f}

In the main text, we show that \ce{TaO2F} is also predicted to have a homochiral $P3_121$ ground state (Section~\ref{sec:extension}). Here we describe the calculations behind this result and give their energies. We calculated DFT energies of \ce{TaO2F} with the settings used for \ce{NbO2F} (Methods), with the \ce{Ta}\_pv projector-augmented-wave potential in place of \ce{Nb}\_sv. Each \ce{TaO2F} arrangement is an arrangement of the \ce{NbO2F} set of Section~\ref{sec:dft} with Ta in place of Nb, so the two compounds can be compared arrangement by arrangement. On the $3\times3\times3$ cell, the set contains the twelve cis + OOF orbits and eight arrangements from each of the other two classes: OOF order with random inter-chain phases, and fully random. For each of these two classes, we ranked the \ce{NbO2F} arrangements by energy and took eight at evenly spaced ranks, including the lowest and the highest.

Table~\ref{tab:si-tao2f} gives the energies of the twelve orbits in both compounds. $P3_121$ is the lowest orbit in \ce{TaO2F}, as in \ce{NbO2F}. The other eleven follow the \ce{NbO2F} order, except that two pairs swap places; the two orbits in each pair differ in energy by \qty{0.02}{\milli\electronvolt\per\atom} or less in \ce{TaO2F}. Arrangements with OOF order but random inter-chain phases lie \qtyrange{4.2}{11.6}{\milli\electronvolt\per\atom} above $P3_121$ (\ce{NbO2F}: \qtyrange{3.8}{12.3}{\milli\electronvolt\per\atom}), and fully random arrangements lie \qtyrange{51}{137}{\milli\electronvolt\per\atom} above $P3_121$ (\ce{NbO2F}: \qtyrange{41}{117}{\milli\electronvolt\per\atom}). The highest orbit lies \qty{1.65}{\milli\electronvolt\per\atom} above $P3_121$, so all twelve orbits lie below every other computed arrangement.

We also calculated DFT energies of \ce{TaO2F} for the fifteen achiral arrangements on the $3\times3\times6$ cell that obey both rules (Section~\ref{sec:si-336}), relative to $P3_121$ in the same cell (Table~\ref{tab:si-336}). All fifteen lie \qtyrange{0.35}{1.59}{\milli\electronvolt\per\atom} above $P3_121$, and the three stackings of $P3_121$ with its mirror image are again the lowest, at \qtyrange{0.35}{0.46}{\milli\electronvolt\per\atom}. Their order is close to that in \ce{NbO2F}: the four pairs that swap places differ in energy by at most \qty{0.11}{\milli\electronvolt\per\atom}.

\begin{table}[tb]
  \caption{\label{tab:si-tao2f} DFT energies of the twelve cis + OOF orbits in \ce{NbO2F} and \ce{TaO2F}, relative to $P3_121$ in each compound. The orbits are in the order of Fig.~\ref{fig:si-gallery}.}
  \begin{ruledtabular}
  \begin{tabular}{lcc}
    & \multicolumn{2}{c}{$\Delta E$ (meV/atom)} \\
    Space group & \ce{NbO2F} & \ce{TaO2F} \\
    \hline
    $P3_121$ & 0 & 0 \\
    $C2$ & 0.52 & 0.38 \\
    $P1$ & 0.86 & 0.74 \\
    $C2$ & 0.92 & 0.76 \\
    $P1$ & 1.25 & 1.00 \\
    $P1$ & 1.40 & 1.10 \\
    $C2$ & 1.46 & 1.08 \\
    $P1$ & 1.54 & 1.24 \\
    $P1$ & 1.73 & 1.34 \\
    $P1$ & 1.76 & 1.39 \\
    $P1$ & 1.82 & 1.38 \\
    $P3_1$ & 2.17 & 1.65 \\
  \end{tabular}
  \end{ruledtabular}
\end{table}

\section{Vacancy-ordered structures}
\label{sec:si-vacancy}

In the main text, we state that in \ce{BaZnO2} and \ce{BaCoO2} the oxygen and vacancies adopt the $P3_121$ arrangement predicted for \ce{NbO2F}, and that in $\beta$-\ce{Fe2SeO} the iron atoms and vacancies adopt the $P3_1$ arrangement (Section~\ref{sec:extension}). Here we describe how we mapped the published structures onto these arrangements.

All three compounds are perovskite-type structures $ABX_3$ in which one third of the $X$ sites are vacant. 
The $B$ sites form a primitive cubic array and the $X$ sites lie midway between neighbouring $B$ sites, so the $B$ and $X$ sites form the same network as the cation and anion sites of \ce{NbO2F}. In \ce{BaZnO2} and \ce{BaCoO2}, $A$ is Ba, $B$ is Zn or Co, and $X$ is O. In $\beta$-\ce{Fe2SeO}, which is an antiperovskite, $A$ is Se, $B$ is O, and $X$ is Fe.

To map each structure onto this parent, we fitted the origin and three lattice vectors of the parent lattice to the positions of the $A$ atoms by least squares. The fitted lattice carries the strain of the crystal relative to the cubic parent, so the remaining offsets of atoms from their ideal sites are local displacements. The ideal $B$ sites lie at the centres of the cells, and the ideal $X$ sites at the centres of the cell faces. We count an ideal $X$ site as occupied if an $X$ atom lies within a quarter of the lattice parameter (about \qty{1}{\angstrom}) of it, and as vacant otherwise. We then labelled each $B$ site by the axis along which both of its neighbouring $X$ sites are occupied, as in Section~\ref{sec:si-enumeration}, and compared the labelling with the twelve orbits under the full symmetry of the $3\times3\times3$ cell. A match with one of the twelve shows that the arrangement obeys both rules and identifies which of the twelve it is.

The refined structures are small distortions of these ideal parent structures. In \ce{BaZnO2} and \ce{BaCoO2}, the three axes of the fitted lattice are equal in length to within 1\%, and the angles between them differ from \ang{90} by at most \ang{2.7}. Every O atom lies within \qty{0.55}{\angstrom} of its ideal site, and the Zn and Co atoms, which sit in tetrahedra, lie within \qty{0.54}{\angstrom} of theirs. In $\beta$-\ce{Fe2SeO}, the fitted lattice is cubic to within 0.1\% in length and \ang{0.2} in angle, every Fe atom lies within \qty{0.51}{\angstrom} of its ideal site, and every O atom lies within \qty{0.38}{\angstrom} of its ideal site. In all three compounds, no atom of the majority species lies within \qty{2.32}{\angstrom} of a vacant site.

\begin{table}[htb]
  \caption{\label{tab:si-vacancy} Published vacancy-ordered structures mapped onto the twelve chiral arrangements. Each structure is listed with the space group reported in its refinement and with the one of the twelve arrangements that its pattern of occupied and vacant sites matches. A pattern and its mirror image count as the same arrangement, so the $P3_2$ refinement of $\beta$-\ce{Fe2SeO} matches the $P3_1$ arrangement.}
  \begin{ruledtabular}
  \begin{tabular}{lll}
    Compound & Space group & Arrangement \\
    \hline
    \ce{BaZnO2}~\cite{vonSchneringEtAl_ZAnorgAllgChem1960} & $P3_121$ & $P3_121$ \\
    \ce{BaCoO2}~\cite{DiattaEtAl_JAmChemSoc2024} & $P3_121$ & $P3_121$ \\
    $\beta$-\ce{Fe2SeO}~\cite{ShamovaEtAl_ChemMater2026} & $P3_2$ & $P3_1$ \\
  \end{tabular}
  \end{ruledtabular}
\end{table}

The refined structures of \ce{BaZnO2} and \ce{BaCoO2} both map onto the $P3_121$ arrangement, and the refinements of $\beta$-\ce{Fe2SeO} at \qtylist{60;110;293}{\kelvin} map onto the $P3_1$ arrangement (Table~\ref{tab:si-vacancy}). The crystal refined by Shamova et al.\ is the left-handed enantiomer, $P3_2$, which belongs to the same orbit~\cite{ShamovaEtAl_ChemMater2026}. In every case, the published space group is that of the ideal arrangement of occupied and vacant sites, so the displacements in the refined structures do not lower its symmetry.

\fi

\end{document}